\UseRawInputEncoding
\documentclass[preprint,12pt, review]{elsarticle}

\journal{...}

\usepackage{amssymb}
\usepackage{amsthm}
\usepackage{graphicx}
\usepackage[version=4]{mhchem}
\usepackage{amsmath}
\usepackage{bm}
\usepackage{hyperref}
\biboptions{sort&compress}

\usepackage{cleveref}

\usepackage{siunitx}
\usepackage{float}
\usepackage{booktabs}
\usepackage{subfig}

\usepackage{bm}
\newcommand{\phim}{\boldsymbol \Phi}
\newcommand{\fm}{\bm{f}}
\DeclareMathAlphabet\mathbfcal{OMS}{cmsy}{b}{n}
\newcommand{\Jac}{\mathbfcal{J}}

\graphicspath{{figs/}}

\newcommand{\tx}[1]{\textrm{#1}}

\usepackage[T1]{fontenc}

\begin{document}

\begin{frontmatter}

\title{Fast reactive flow simulations using analytical Jacobian and dynamic load balancing in OpenFOAM}

\author[1]{Ilya Morev$^*$}
\author[1]{Bulut Tekg\"ul}
\author[1]{Mahmoud Gadalla}
\author[1]{Ali Shahanaghi}
\author[1]{Jeevananthan Kannan}
\author[1]{Shervin Karimkashi}
\author[1]{Ossi Kaario}
\author[1]{Ville Vuorinen}

\address[1]{Department of Mechanical Engineering, Aalto University School of Engineering, Otakaari 4, 02150 Espoo, Finland.}

\cortext[*]{Corresponding Author. E-mail address: ilya.morev@aalto.fi}

\begin{abstract}
Detailed chemistry-based computational fluid dynamics (CFD) simulations are computationally expensive due to the solution of the underlying chemical kinetics system of ordinary differential equations (ODEs). Here, we introduce a novel open-source library aiming at speeding up such reactive flow simulations using OpenFOAM, an open-source software for CFD. First, our dynamic load balancing model by Tekg\"ul et al.~[DLBFoam:  An open-source dynamic load balancing model for fast reacting flow simulations in OpenFOAM, Computer Physics Communications 267 (2021)] is utilized to mitigate the computational imbalance due to chemistry solution in multiprocessor reactive flow simulations. Then, the individual (cell-based) chemistry solutions are optimized by implementing an analytical Jacobian formulation using the open-source library \texttt{pyJac}, and by increasing the efficiency of the ODE solvers by utilizing the standard linear algebra package (\texttt{LAPACK}). We demonstrate the speed-up capabilities of this new library on various combustion problems. These test problems include a 2-dimensional (2D) turbulent reacting shear layer and 3-dimensional (3D) stratified combustion to highlight the favourable scaling aspects of the library on ignition and flame front initiation setups for dual-fuel combustion. Furthermore, two fundamental 3D demonstrations are provided on non-premixed and partially premixed flames, namely the Engine Combustion Network Spray A and the Sandia flame D experimental configurations, which were previously considered unfeasible using OpenFOAM. The novel model offers up to two orders of magnitude speed-up for most of the investigated cases. The openly shared code along with the test case setups represent a radically new enabler for reactive flow simulations in the OpenFOAM framework.
\end{abstract}


\end{frontmatter}


\section{Introduction}
\label{sec:intro}
There is an urgent need to develop effective and accurate engineering solutions to combat climate change. Reactive flow simulations play an essential role in academic and industrial research on mitigating harmful emissions. A number of computational fluid dynamics (CFD) packages are available on the market, offering a variety of features together with possible limitations. Major limitations related to software distribution under commercial licenses are their affordability and source code access. The imminent pace of climate change urges us to broaden the use of the latest achievements in CFD beyond the limitations of commercial software to all appropriate applications. Unfortunately, the available open-source reacting flow solvers are currently limited. The most popular open-source general-purpose CFD code -- OpenFOAM -- poses relatively poor performance for finite rate chemistry simulations compared to commercial solutions, making it impractical for large-scale industrial applications and thus limiting its use to relatively simple academic cases. However, the importance of open-source solutions was recently recognized at a governmental level. For instance, there is a large ongoing project funded by the European Union on optimizing the OpenFOAM code -- exaFoam~\cite{exaFoam}. In this work, we proceed in a similar track and focus on the search for the bottlenecks in reacting flow simulations within OpenFOAM with the aim of significant computational speed-up.

One of the most fundamental approaches for reactive flow modeling is to use CFD with the direct integration of chemical kinetics, referred to as finite-rate chemistry~\cite{Poinsot2011theoretical}.
In this approach, each computational cell is treated as an individual chemistry problem with pressure~($p$), and a thermochemical state vector~($\phim$) comprising temperature~($T$) and species mass fractions~($Y_k$). The rate of change of the thermochemical state vector, $\partial \phim /\partial t = \fm (\phim, t)$, within a computational cell forms stiff system of ordinary differential equations (ODEs) that requires a specific class of algorithms for temporal integration.

In reactive CFD applications, chemistry evaluation comprises the most computationally demanding part of the simulation. With the growing complexity of chemical kinetic mechanisms, the cost of solving the chemistry problem often exceeds the cost of fluid flow solution by up to two orders of magnitude~\cite{Peters2000}. Nevertheless, high-fidelity reactive flow simulations can be still feasible through efficient usage of the provided computational resources, in addition to cell-based optimization of the chemistry solution~\cite{Tekgul2020dlbfoam}. In general,  considerable efforts have been made in the literature to improve the computational performance of CFD simulations involving turbulence~\cite{Kochkov2021,Gadalla2021a,Hijazi2020, Wandel2021, Li2019}, shock waves~\cite{Brahmachary2021}, gas rarefaction~\cite{Plimpton2019}, geometrical parameterization~\cite{Sekar2019}, or chemical reactions~\cite{Aversano2021,Wehrfritz2016, Jaganath2021,Contino2011} through model reduction, machine learning, or efficient parallelization.

In finite-rate chemistry simulations, the high computational cost associated with the chemistry solution originates mainly from two factors. First, the cost of solving the system of ODEs depends on the system stiffness which is, to some extent, influenced by the local thermochemical state. For example, an oxidation process comprising different radicals at various chemical timescales may constitute a greater computational challenge compared to a local state of burnt mixture. Second, the cost of computing the rate of change of the thermochemical state vector is directly proportional to the chemical mechanism size~\cite{Lu2009, Law2007}, which can reach up to thousands of species and even more reactions~\cite{Pei2015, Sarathy2016}. 

For the chemistry ODE problem, implicit solvers involving Jacobian evaluation are usually preferred over explicit ones for stability and accuracy reasons~\cite{Hairer1996}. The evaluation of the Jacobian, $\Jac = \partial \fm / \partial \phim $, is a computationally demanding process within an ODE solution routine. ODE solvers often employ finite differencing methods for its evaluation, which may be an expensive operation depending on the chemical mechanism size. Lu et al.~\cite{Lu2009} reported that the computational cost of numerical evaluation of Jacobian scales quadratically with the number of reactions. It has been shown that using an analytically computed Jacobian provides high performance gain during the ODE system solution~\cite{Lu2009,McNenly2015,Perini2012}. There are various implementations of the analytical Jacobian matrix evaluation in the literature~\cite{Schwer2002,Safta2011,Perini2012}. Recently, Niemeyer et al.~\cite{Niemeyer2017} introduced an open-source library \texttt{pyJac}, which generates C subroutines for analytical Jacobian evaluation. They demonstrated its accuracy and out-performance over the available analytical Jacobian generators in the literature.

While the cost of integrating a single chemistry ODE problem poses a performance challenge in reactive CFD simulations, there is a secondary issue which is often left unaddressed. As previously mentioned, the computational cost of solving a chemistry ODE depends on the local stiffness of $\phim$. The difference in computational complexity throughout the geometrical domain poses a load imbalance issue for multiprocessor applications, where one process becomes the computational bottleneck and causes performance issues. There have been several studies in the recent literature tackling this load imbalance issue via utilizing dynamic load balancing algorithms, often using Message-Passing Interface (MPI) routines. Antonelli et al.~\cite{Antonelli2011} developed an MPI-based model introducing a cell distribution based load balancing algorithm. Following this work, Shi et al.~\cite{Shi2012} and Kodavasal et al.~\cite{Kodavasal2016} both introduced stiffness-detection based load balancing algorithms and employed them in reactive CFD simulations. Zirwes et al. developed an MPI-based dynamic load balancing algorithm for chemistry problem distribution for OpenFOAM~\cite{Zirwes2018optimizing}. More recently, Muela et al.~\cite{Muela2019} presented a dynamic load balancing model, which also utilizes a stiffness detection algorithm that chooses the optimal ODE integration method for each computational cell. These methods introduced computational speed-up around \numrange{3}{5}, depending on the application. We have recently also introduced an open-source dynamic load balancing model called \texttt{DLBFoam} for OpenFOAM, providing speed-up up to a factor of $10$~\cite{Tekgul2020dlbfoam}.

In this work, we introduce a novel chemistry model that provides speed-up in reactive flow simulations in OpenFOAM by targeting the two major issues in reactive CFD simulations addressed above. We further extend our dynamic load balancing model \texttt{DLBFoam}, and focus on the optimization of the cell-wise chemistry solution by introducing two new features. First, we utilize an analytical Jacobian approach using the \texttt{pyJac} library. Second, we make use of the standard linear algebra library (LAPACK)~\cite{LAPACK1999} to further improve the ODE solution procedure of the chemistry problem, by 
replacing the LU decomposition and back substitution operations of the Seulex ODE integration algorithm in OpenFOAM with more robust alternatives suitable for dense Jacobian matrices. The effectiveness and robustness of the developed model are demonstrated on two different academic dual-fuel (DF) combustion set-ups: a 2-dimensional (2D) reacting shear layer and a 3-dimensional (3D) stratified combustion configuration. After that, the model is further demonstrated in two well-established experimental flame configurations: a non-premixed $n$-dodecane flame (Engine Combustion Network (ECN) Spray A), and a partially-premixed methane-air flame (Sandia flame D).

The paper is outlined as follows. First, implementation details of the developed model are presented in \Cref{sec:implementation}. Then, in \Cref{sec:showcases} we discuss the application of our model to different combustion simulations, represented by the aforementioned test cases. Finally, conclusions are provided in \Cref{sec:conc}. 

\section{Implementation}
\label{sec:implementation}
In this section, implementation details of our model are presented. We note that an earlier in-house version of this implementation was carried out by Kahila et al.~\cite{Kahila2019}, details of which can be found in his thesis work~\cite{Kahila2019Thesis}. This section briefly discusses the \texttt{DLBFoam} model introduced by Tekg\"ul et al.~\cite{Tekgul2020dlbfoam} as an improvement to our in-house code, and further describes the efforts towards optimizing the solution of the chemistry ODE problem by utilizing an analytical Jacobian formulation (\texttt{pyJac}) together with efficient linear algebraic routines (\texttt{LAPACK}).

\subsection{Finite rate chemistry}
\label{section:finiteref}

In reactive CFD applications, we numerically solve the governing equations for mass, momentum, energy, and species transport. The total production or consumption rate of chemical species, and the heat release rate are represented as source terms in the species transport equations and the energy equation, respectively. Due to the vast difference between the chemistry and flow time scales, commonly an operator splitting procedure is utilized to separate the calculation of source terms from the flow solution. These source terms define the change of the local thermochemical state which results from solving a stiff, nonlinear initial value problem described as:
\begin{equation}
\label{eq1}
\centering
\begin{cases}
\partial \phim / \partial t =  \left\{\frac{\partial T}{\partial t}, \frac{\partial Y_1}{\partial t}, \frac{\partial Y_2}{\partial t},..., \frac{\partial Y_{N_{sp}-1}}{\partial t}\right\}^\intercal = \fm(\phim,t) , \\
\phim(t=0) = \phim_0 ,
\end{cases}
\end{equation}
where $N_{sp}$ is the number of species in the chemical mechanism. The ODE system in \Cref{eq1} is integrated in each computational cell over the CFD time step $\Delta t_{\text{CFD}}$ in smaller chemical time steps $\Delta t_{\text{ODE}} \leq \Delta t_{\text{CFD}}$ (see Part C for more information). The source terms are then evaluated and used in species transport and energy equations. Further details on this modeling approach can be found in the work of Imren and Haworth~\cite{Imren2016}.

\subsection{Dynamic load balancing: DLBFoam}
Next, the previously developed \texttt{DLBFoam} library is briefly introduced here as essential background information. As mentioned earlier, the computational cost of solving the stiff nonlinear ODE may depend on the local thermochemical conditions. In multiprocessor reactive CFD simulations, the process with the highest computational load may take longer to compute than the rest, hence creating a bottleneck within the given CFD time step.

\begin{figure}[H]
	\centering
    {\includegraphics[width=0.44\columnwidth]{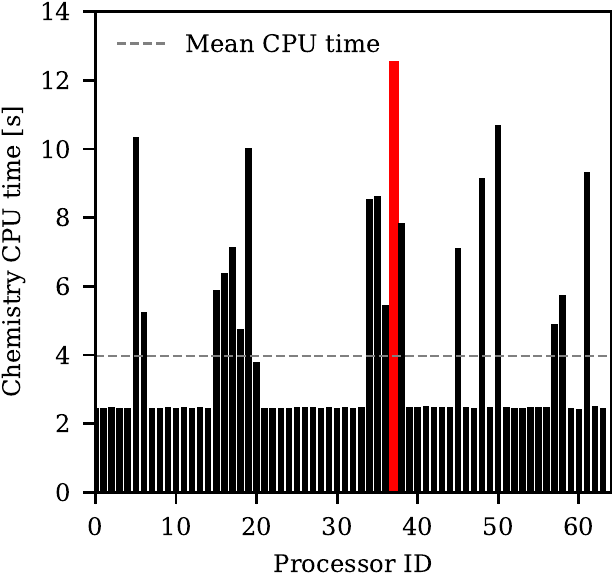} }
    \
    {\includegraphics[width=0.44\columnwidth]{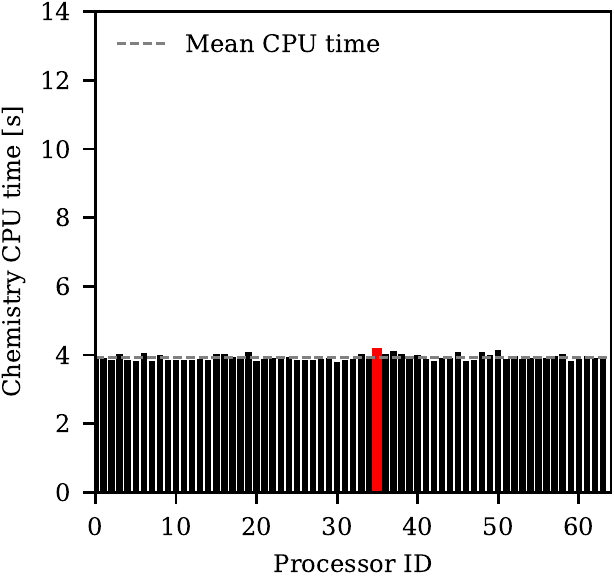} }
	\caption{Left: The process-based computational imbalance in reactive CFD simulations without \texttt{DLBFoam}. The process with the highest computational load (red) creates a bottleneck in the simulation. Right: Same case with \texttt{DLBFoam}.}
	\label{fig:imbalance}
\end{figure}

Recently, we introduced an open-source model called \texttt{DLBFoam}~\cite{Tekgul2020dlbfoam} for OpenFOAM, aiming to mitigate this imbalance issue in multiprocessor reactive CFD simulations via dynamic load balancing. \texttt{DLBFoam} uses MPI routines to redistribute the chemistry computational load evenly between processes during the simulation. \Cref{fig:imbalance} demonstrates the computational imbalance caused by the direct integration of chemistry, and how it is mitigated by \texttt{DLBFoam}. In addition, \texttt{DLBFoam} introduces a zonal reference cell mapping method, which further lowers the computational cost by mapping a chemistry solution from a reference cell instead of explicitly solving it for ambient regions with low reactivity. In total, we reported around a factor of $10$ speed-up in 3D reactive CFD simulations, compared to the standard OpenFOAM chemistry model. Further implementation details on \texttt{DLBFoam} can be found in our previous publication describing it in detail~\cite{Tekgul2020dlbfoam}.

\subsection{ODE solver optimization and analytical Jacobian}
\label{sec:ode_opt}

The new development in the present paper introduces a coupling of 1) the analytical Jacobian formulation introduced in \texttt{pyJac} library, 2) standard linear algebra routines from \texttt{LAPACK} library for the ODE solution, and 3) the \texttt{DLBFoam} library previously introduced by Tekg\"{u}l et al.~\cite{Tekgul2020dlbfoam}. Within a CFD time step, chemistry is treated as an independent, stiff system of ODEs. Implicit and semi-implicit algorithms for solving this system of stiff ODEs are usually preferred to ensure the solution stability~\cite{Hairer1996}.

The solution to the nonlinear initial value problem \Cref{eq1}, can be obtained using a variant of Newton iteration with the following discretized form  
\begin{equation}
\label{eq2}
\begin{split}
    \phim_{n+1} &= \phim_n + \int_{t_n}^{t_n+\Delta t_{\text{ODE}}} \left( \fm_n + \Jac_n (\phim_{n+1} - \phim_n) + \mathcal{O} (\Delta t_{\text{ODE}}^2) \right) dt \\
    &= \phim_n + \fm_n \Delta t_{\text{ODE}} + \Jac_n (\phim_{n+1} - \phim_n) \Delta t_{\text{ODE}} + \mathcal{O} (\Delta t_{\text{ODE}}^2),
\end{split}
\end{equation}
which is then linearized by neglecting the higher-order terms. Here, $\Delta t_{\text{ODE}}$ is a subinterval of the full integration interval $\Delta t_{\text{CFD}}$, which is commonly determined by the underlying ODE solver using a combination of theoretical and heuristic relations. The direct solution of the previous equation for $\phim_{n+1}$ requires the inverse of $\Jac$, which is commonly avoided through matrix factorization (i.e., LU decomposition) and back substitution techniques. 

Commonly in reactive CFD codes, the Jacobian of the system of ODEs representing chemistry is implemented via finite differencing to be used in ODE solver integration. However, using such an approximation introduces a negative effect on the ODE solver accuracy and performance~\cite{Niemeyer2017}. Furthermore, calculating the Jacobian with finite differencing is a computationally expensive procedure~\cite{Lu2009,Niemeyer2017}. Instead, utilizing a fully algebraic Jacobian is more efficient and it improves the ODE solver accuracy and performance. Lu et al.~\cite{Lu2009} reported that using an analytical Jacobian reduces its evaluation time from square of the number of species to a linear dependence.

\begin{figure}[H]
    \centering
    \includegraphics[width=0.95\linewidth]{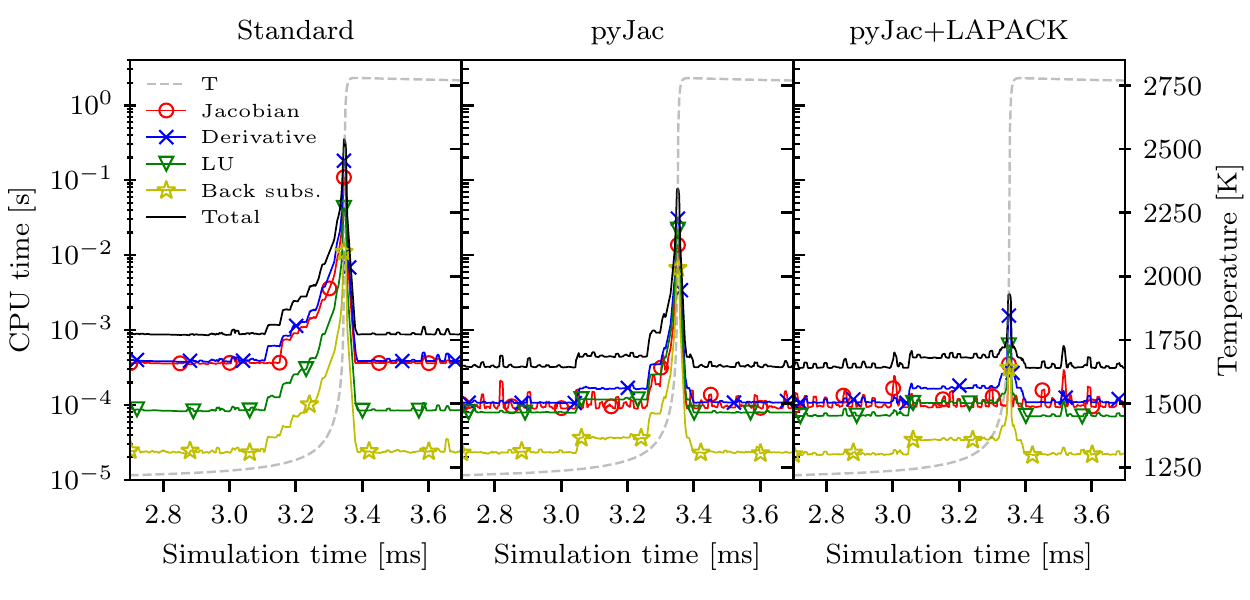}
    \caption{CPU time spent on different operations of ODE solution algorithm for a single cell 0D problem involving \ce{CH4} chemistry using \texttt{Standard}, \texttt{pyJac} and \texttt{pyJac+LAPACK} models. Selected timeframe represents the most computationally demanding part of the problem. The GRI-3.0 mechanism ($53$ species and $325$ reactions) is used in the benchmark.}
    \label{fig:seulex_timed} 
\end{figure}

OpenFOAM features a fully algebraic Jacobian implementation to speed up the solution of chemical kinetics in its recent releases~\cite{of_jacobian}. However, we have found that the implemented algebraic Jacobian fails to deliver a fast solution to chemistry ODE integration when high ODE convergence tolerances are utilized. Instead, we utilize the python-based open-source software \texttt{pyJac}, introduced by Niemeyer et al.~\cite{Niemeyer2017}. The \texttt{pyJac} package generates C subroutines for evaluation of an analytical Jacobian for a chemical mechanism. They reported that \texttt{pyJac} performs \numrange{3}{7.5} times faster than the first-order finite differencing approach. In our proposed model, OpenFOAM's Jacobian calculation subroutines are replaced with the subroutines generated by \texttt{pyJac}.

The fully algebraic Jacobian utilization via \texttt{pyJac} speeds up the Jacobian evaluation time and improves the solution convergence rate. However, the ODE solution method itself is also particularly important for obtaining faster chemistry solution. We observed that the OpenFOAM's native functions used for LU decomposition and back substitution matrix operations of the chemistry ODE system are not suitable for solving dense and small matrices (\num{< 500 x 500}).  Therefore the LU decomposition and back substitution linear algebraic operations of OpenFOAM are replaced by the more robust implementations existing in the open-source library \texttt{LAPACK}~\cite{LAPACK1999}. The \texttt{LAPACK} routines utilized in our model are very efficient for dense and small matrices, such as the fully dense Jacobian matrix created by the \texttt{pyJac}.

In this study, \texttt{LAPACK} functions were linked to a semi-implicit extrapolation-based Euler method, denoted as Seulex~\cite{Hairer1996,Imren2016}. Seulex requires the specification of relative and absolute tolerances of the solution along with an initial estimate of $\Delta t_{\text{ODE}}$. As the solution progresses, the solver improves the estimate for $\Delta t_{\text{ODE}}$ at the current thermochemical state using theoretical relations and heuristics.  The previous available value is then used as an initial value for the following integration interval. Furthermore, a $\Delta t_{\text{ODE}}$ is recursively split into a number of partitions and each partition is then solved using a low-order method. Finally, the high-order solution over the whole interval $\Delta t_{\text{ODE}}$ is recursively gathered from solutions of the smaller partitions. If the solution does not satisfy the tolerance criteria, partitioning continues until it is satisfied. A more detailed explanation of the Seulex algorithm can be found in the work by Imren and Haworth~\cite{Imren2016}.

As noted earlier, each computational cell has its own independent ODE problem with a thermochemical state vector comprising only intensive mixture properties. Furthermore, the ODE solver guarantees that if the solution has converged, the error will lie within user-specified tolerance regardless of the width of the integration interval $\Delta t_{\text{CFD}}$. However, the error related to the operator splitting technique needs to be taken care of. The assumption behind the operator splitting is that the chemistry timescales are orders of magnitude smaller than the flow timescales or the CFD solver time step $\Delta t_{\text{CFD}}$. Nevertheless, the time step $\Delta t_{\text{CFD}}$ is to be chosen so that the species and temperature transport effects over a time step are accounted for and the thermochemical state vector is updated correspondingly. We note that in this study we do not consider any chemical source term closures for the filtered or averaged equations. Such  turbulence-chemistry interaction (TCI) models could significantly depend on the mesh parameters and the CFD time step size and might require a separate treatment.

The effect of the optimized Jacobian evaluation is demonstrated in \Cref{fig:seulex_timed} on a 0-dimensional (0D) homogeneous reactor simulation with stoichiometric methane-air mixture at $T = \SI{1200}{K}$ and $p = \SI{13.5}{atm}$. The GRI-3.0~\cite{gri} chemical kinetic mechanism was used and the absolute and relative ODE solver tolerances were set to $10^{-10}$ and $10^{-6}$, respectively. The integration interval ($\Delta t_{\text{CFD}}$) was fixed to $10^{-6}~\si{s}$ which corresponds to timescales relevant to reactive CFD applications under engine-relevant conditions. From the figure, it is observed that the time spent on Jacobian and derivative evaluations is reduced by roughly one order of magnitude between \texttt{Standard} and \texttt{pyJac} along the entire simulation time. Furthermore, the exact Jacobian formulation increases the solution accuracy for every Newton iteration, which reduces the amount of iterations required for convergence. Next, with \texttt{LAPACK} we notice a further reduction in the analytical Jacobian retrieval and also less time for all operations only within the sharp gradient zone. The reason is that with \texttt{LAPACK} the ODE solver uses wider sub-interval in the stiff zone, hence, fewer function calls for the Jacobian retrieval and consequently, fewer linear algebraic operations.

 \begin{figure}[H]
    \centering
    \includegraphics[width=85mm]{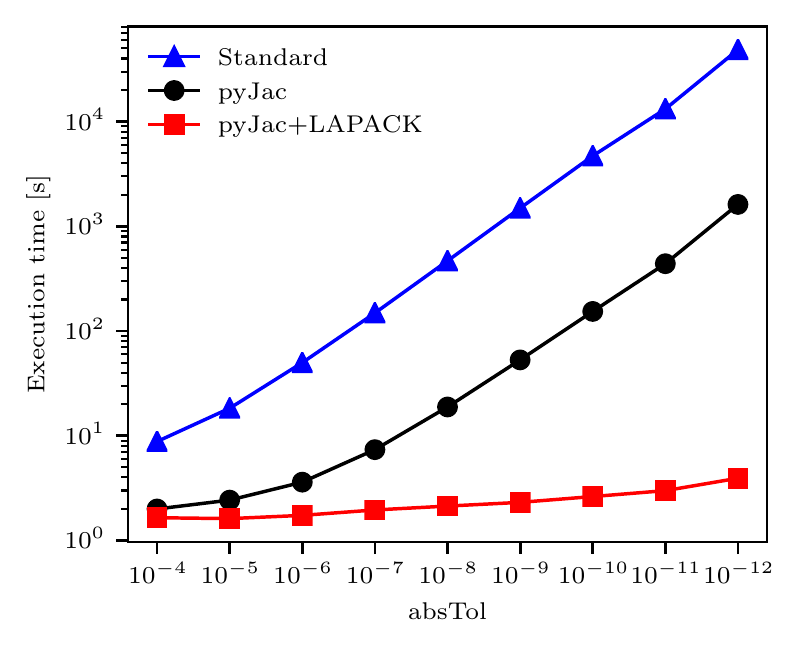}
    \caption{Dependency of the execution time on the absolute tolerance ($N_{sample}=10$) for a single cell 0D problem involving \ce{CH4} chemistry with different ODE convergence tolerances. The GRI-3.0 mechanism ($53$ species and $325$ reactions) is used in the benchmark.}
    \label{fig:0d}
\end{figure}

\Cref{fig:0d} depicts the total execution times for the aforementioned problem with varied absolute tolerances.
It can be seen that while the Jacobian retrieval using \texttt{pyJac} reduces the execution time by about one order of magnitude, using \texttt{LAPACK} routines provides another order of magnitude speed-up. Utilization of \texttt{pyJac} together with \texttt{LAPACK} allows using much tighter ODE tolerances.

\section{Results}
\label{sec:showcases}

\subsection{An overview of the test cases}

\begin{figure}[H]
    \centering
    \subfloat[
    ] {{\includegraphics[width=0.45\columnwidth]{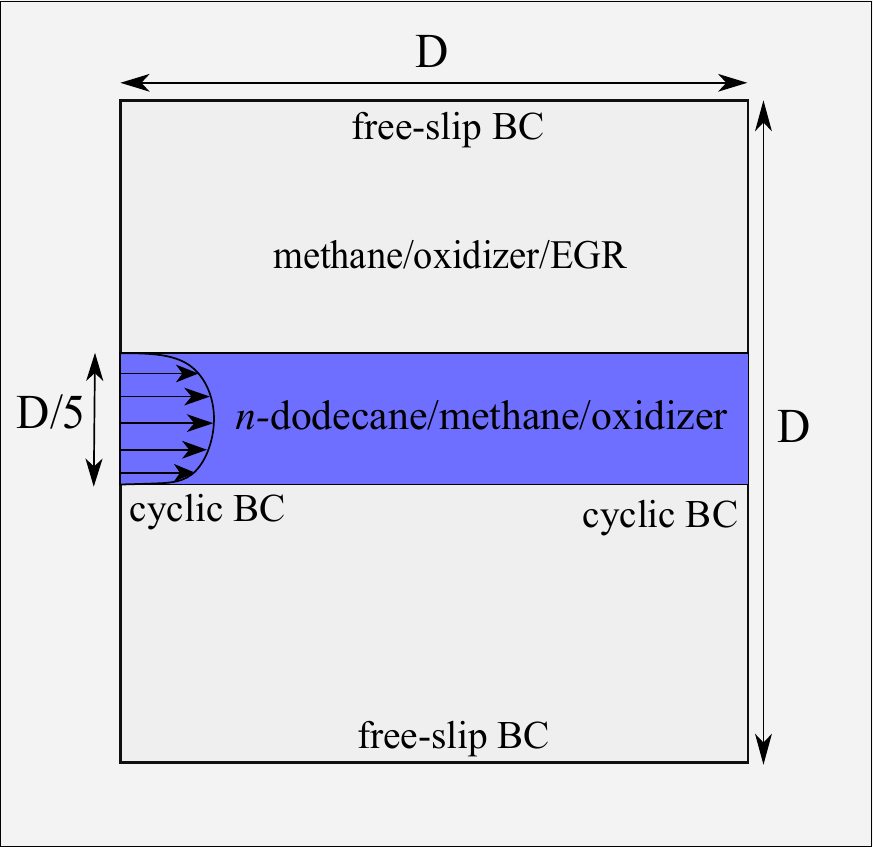} }\label{fig:schematic_2dshear}} \quad
    \subfloat[
    ]{{\includegraphics[width=0.45\columnwidth]{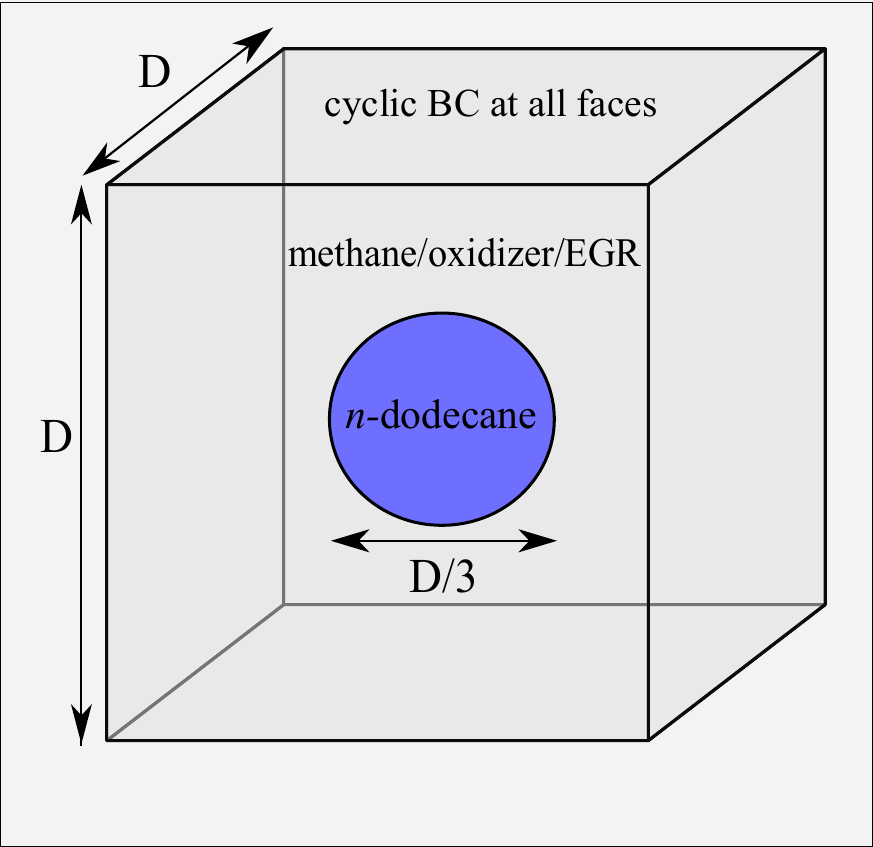} }\label{fig:schematic_3dcomb}} \\[2ex]
    \subfloat[
    ] {{\includegraphics[width=0.45\columnwidth]{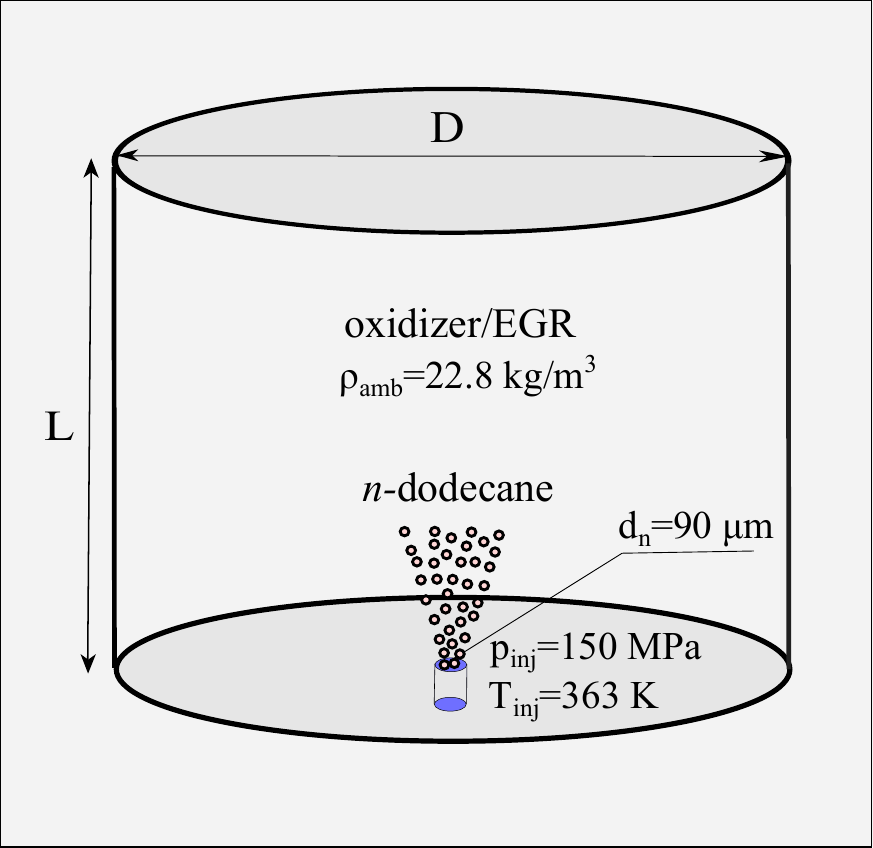} }\label{fig:schematic_sprayA}} \quad
    \subfloat[
    ]          {{\includegraphics[width=0.45\columnwidth]{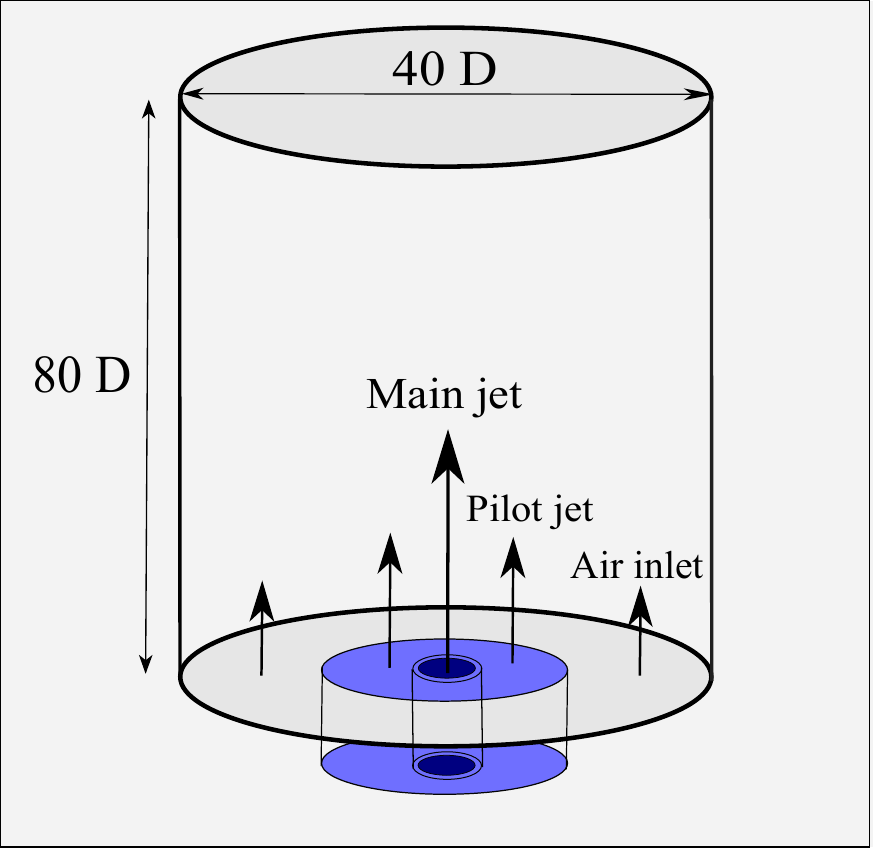} }\label{fig:schematic_sandiaD}}%

	\caption{Schematic diagrams of the demonstrated test cases. (a) 2D reacting shear layer, (b) 3D stratified combustion, (c) ECN Spray A, (d) Sandia flame D.
	}
	\label{fig:schematic_diagrams}
\end{figure}

After the introduction of the new features of our model in the previous section, here the combined effect of the \texttt{DLBFoam} library with \texttt{pyJac} and \texttt{LAPACK} is investigated. Three different models are employed here for performance benchmarking: 1) the standard OpenFOAM chemistry model (referred to as \texttt{Standard}), 2) our original model with dynamic load balancing and zonal reference mapping~\cite{Tekgul2020dlbfoam} (referred to as \texttt{DLBFoam}), and 3) our original model combined with the ODE related improvements utilizing \texttt{pyJac} and \texttt{LAPACK} routines (referred to as \texttt{DLBFoam+pyJac}). We first demonstrate the computational performance of \texttt{DLBFoam+pyJac} in two academic test cases (2D reacting shear layer and 3D stratified combustion). After that, we demonstrate its performance in two experimental flame configurations (ECN Spray A and Sandia flame D). Schematic diagrams presenting the four test cases are provided in \Cref{fig:schematic_diagrams}.

For each case, we first use \texttt{DLBFoam+pyJac} to solve a full-scale combustion problem. Then, for each problem we choose a time interval which is considered computationally challenging, e.g., chemistry ODE problem is stiff in parts of the domain due to ignition, and compare the performance of \texttt{DLBFoam+pyJac} against \texttt{DLBFoam} and \texttt{Standard} for a specific number of iterations. A summary of various case-specific benchmark parameters along with computational speed-up are presented in \Cref{table:summary}. Here, we have to limit the number of iterations for benchmarking because of the poor performance of the \texttt{Standard} model we highlighted in the previous section. In fact, with the chosen strict tolerance values, it is infeasible to carry out a full simulation with \texttt{Standard} or even benchmark with a larger number of iterations than those stated in the table for all of the studied test cases. 

\begin{table}[H]
    \centering
    \caption{List of parameters used in performance benchmarks. Computational time refers to the full-scale simulation using \texttt{DLBFoam+pyJac}. Computational speed-up is reported for \texttt{DLBFoam+pyJac} against \texttt{Standard}. Computational time for ECN Spray A reports the case with $T_{amb}=\SI{900}{K}$ and simulation time of \SI{1.5}{ms}.} 
    \label{table:summary}
    \small
    \resizebox{1\columnwidth}{!}{%
    \begin{tabular}{ l@{\hskip 4mm}l@{\hskip 3mm}l@{\hskip 3mm}l@{\hskip 3mm}l }
    \toprule
                            & \textbf{2D shear layer} 
                            & \begin{tabular}[c]{@{}c@{}}\textbf{3D stratified}\\\textbf{combustion}\end{tabular}   
                            & \textbf{ECN Spray A} 
                            & \textbf{Sandia flame D} \\
\midrule
\textbf{Chemical mechanism}         & Yao \cite{Yao2017} & Yao       & Yao         & DRM19 \cite{DRM19} \\
\textbf{Number of cells}            & $90\,000$          & $16.78$M  & $39$M       & $1.99$M   \\
\textbf{Number of processors}       & $8$--$32$          & $1152$    & $1920$      & $768$     \\
\textbf{Benchmarking samples (CFD iterations)} & $100$   & $7$       & $65$        & $100$     \\
\textbf{Computational speed-up}     & $24$--$38$         & $\approx 400$     & $256$       & $13.5$     \\
\textbf{Computational time [CPU-hr]} & $65$              & $15\,000$ & $128\,500$  & $24\,500$ \\
\bottomrule
    \end{tabular}
    } 
\end{table}

Regarding numerical schemes, a second-order spatial and temporal discretization is utilized for all test cases presented herein. The reacting PISO (Pressure-Implicit with Splitting of Operators) algorithm with two outer-loop correctors is utilized for pressure-velocity coupling~\cite{Issa1991}, i.e., chemistry is solved twice within a single CFD time step. The absolute and relative tolerances of the chemistry ODE solver are set to $10^{-10}$ and $10^{-6}$, respectively. More details about the case setups can be found in test case files which are openly shared~\cite{DLBFoam_cases_repo}.

All simulations and benchmarks were performed on the Mahti supercomputer at CSC - Finnish IT Center for Science. Mahti has $1404$ nodes each with two $64$-core \SI{2.6}{GHz} (boost up to \SI{3.3}{GHz}) AMD Rome 7H12 CPUs. All nodes are connected to the inter-node communication network with \SI{200}{GB/s} links~\cite{mahti}.

\subsection{2D reacting shear layer}

The first test case presents a simple, temporally evolving 2D reacting shear layer which is computationally affordable even on a personal computer. The presented test case is fundamentally related to the ignition and flame initiation of DF combustion in compression-ignition engines such as the reactivity controlled compression ignition (RCCI) engine~\cite{Kahila2019,Kahila2019a}. We demonstrate the computational speed-up and scaling effects of \texttt{DLBFoam+pyJac} with respect to the number of cores. 

The schematic of the present numerical setup is illustrated in \Cref{fig:schematic_2dshear}. The setup contains a high reactivity $n$-dodecane jet stream that mixes with the surrounding oxidizer consisting of premixed methane, oxidizer and exhaust gas recirculation (EGR). The simulation domain is a square box with a side length $D=\SI{1.5}{\milli\meter}$. The number of grid points in both directions is $300$ which is based on the pre-estimated laminar flame thickness for $n$-dodecane-methane premixed flame at the corresponding most reactive mixture fraction, $p=\SI{60}{bar}$, and $T_\tx{reactants}=\SI{800}{K}$ ($\delta_f \approx \SI{50}{\micro m}$) to resolve the flame by $10$ grid points. A hyperbolic tangent function is used to generate the shear layer between the $n$-dodecane and oxidizer streams. The $n$-dodecane jet is initially set to \SI{700}{K} whereas the methane/air mixture is set to \SI{900}{K} at a constant pressure of \SI{60}{bar}. The initial conditions corresponding to the ambient premixed mixture are similar to our previous DF spray studies~\cite{Kahila2019a,Kahila2019,Tekgul2020,Tekgul2021a, Kannan2020, Kannan2021} as reported in \Cref{tab:composition_shear_layer}. The $n$-dodecane jet moves initially with a relative velocity of \SI{10}{m/s} to the methane/air stream and develops a Kelvin-Helmholtz instability. A skeletal chemical mechanism ($54$ species and $269$ reactions) developed by Yao et al.~\cite{Yao2017} for $n$-dodecane combustion is used. The performance of this mechanism in DF context has been already demonstrated in our earlier studies~\cite{Kahila2019, Tekgul2020,Kannan2020}.  

\begin{table}[H]
    \caption{Initial conditions in molar fractions and corresponding equivalence ratio for premixed ambient mixture in 2D shear layer and 3D stratified combustion cases.} 
    \label{tab:composition_shear_layer}
    \centering
        \begin{tabular}{c c c c c c}
        \hline
             $\ce{CH4}$ \% & $\ce{O2}$ \% & $\ce{CO2}$ \% & $\ce{H2O}$ \% & $\ce{N2}$ \% & $\phi_{\ce{CH4}}$ \\ \hline
             3.750 & 15 & 5.955 & 3.460 & 71.835 & 0.5 \\
        \hline
        \end{tabular}
\end{table}

\begin{figure}[H]
    \centering
    \includegraphics[width=0.99\textwidth]{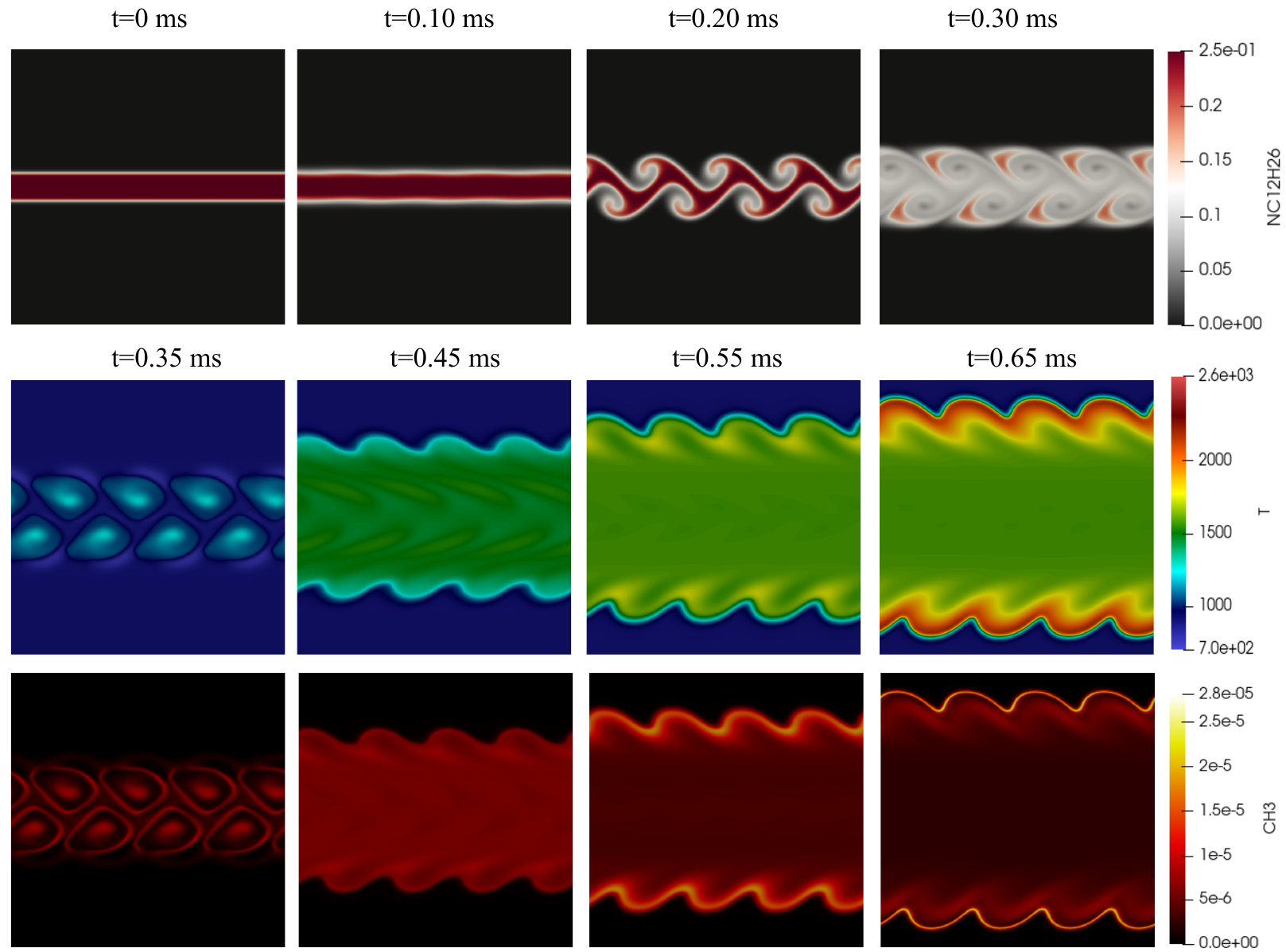}
    \caption{Temporal evolution of $n$-dodecane (top row), temperature (middle row), and \ce{CH3} (bottom row) fields of the 2D reacting shear layer simulation, where the $n$-dodecane jet initiates the flame and then burns the premixed ambient.}
    \label{fig:shear_layer_cutplanes}
\end{figure}

\Cref{fig:shear_layer_cutplanes} highlights the DF ignition process in the shear layer, where $n$-dodecane jet ignites the surrounding mixture. It is observed that the first-stage ignition from low-temperature chemistry develops primarily within the $n$-dodecane jet near the mixing layer. Next, the ignition front initiates in the $n$-dodecane region at the most reactive mixture fraction and propagates towards the ambient mixture. Here, the mixture fraction describes the mixing extent of $n$-dodecane fuel stream and the premixed methane, oxidizer and EGR. Finally, a premixed flame front is established which completes the combustion of the ambient methane. 

\begin{figure}[H]
    \centering
    {{\includegraphics[width=0.47\columnwidth]{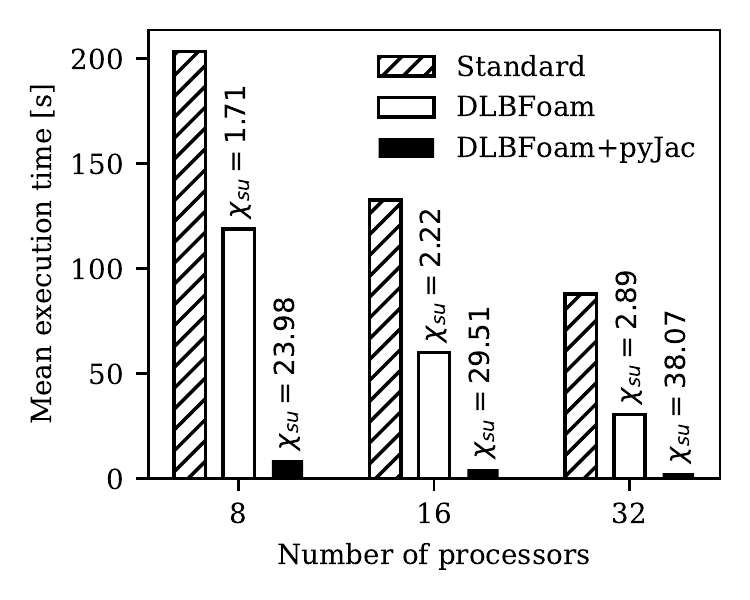} }} \
    {{\includegraphics[width=0.47\columnwidth]{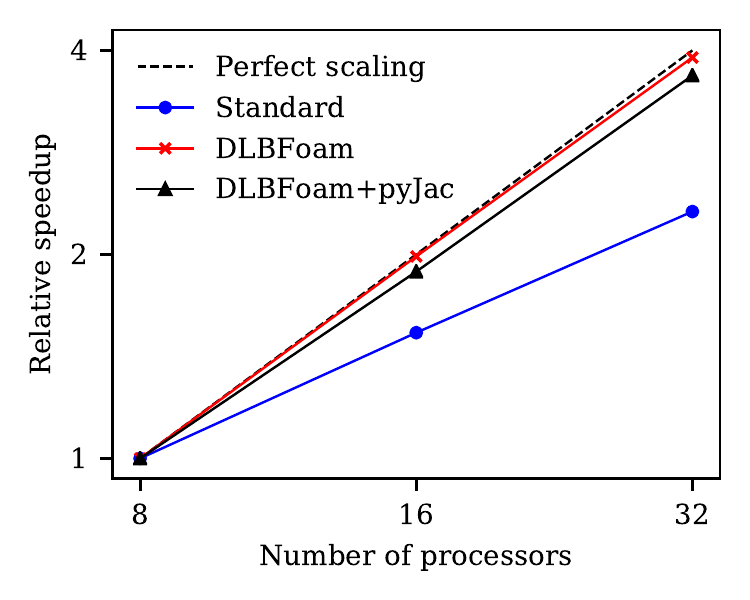} }}%
    
	\caption{Left: Mean execution times over $100$ iterations for the 2D reacting shear layer for different number of processors. Here, $\chi_{su}$ is the speed-up factor of \texttt{DLBFoam} or \texttt{DLBFoam+pyJac} compared to \texttt{Standard}. Right: Speed-up values for different number of processors relative to $8$ processors case.}
	\label{fig:shearlayer_speedup_figure}
\end{figure}

Next, we report the computational speed-up and the parallel efficiency provided by our model for different processor counts, as depicted in \Cref{fig:shearlayer_speedup_figure}. The domain is decomposed into $8$, $16$, and $32$ processors to demonstrate the scaling capabilities of the model. All speed-up tests are carried for $100$ constant CFD time steps of \SI{2e-7}{s} after the DF ignition. The following observations are made from the analysis: 1) the \texttt{Standard} model demonstrates a poor parallel scaling efficiency due to the load imbalance. While the strong scaling efficiency of \texttt{DLBFoam} and \texttt{DLBFoam+pyJac} are almost linear ($>95\%$), the \texttt{Standard} shows a scaling efficiency around $75\%$ for the investigated processor counts. This explains the increased speed-up value for \texttt{DLBFoam} and \texttt{DLBFoam+pyJac} for higher processor counts. 2) The \texttt{DLBFoam} model provides speed-up by a factor of $1.71$ to $2.89$ for different processor counts, compared to \texttt{Standard}. 3) The \texttt{DLBFoam+pyJac} model provides speed-up by a factor of $23.98$ to $38.07$ compared to \texttt{Standard}.

\subsection{3D stratified combustion}
\label{sec:sparks}

\begin{figure}[H]
    \centering
    \includegraphics[width=0.99\textwidth]{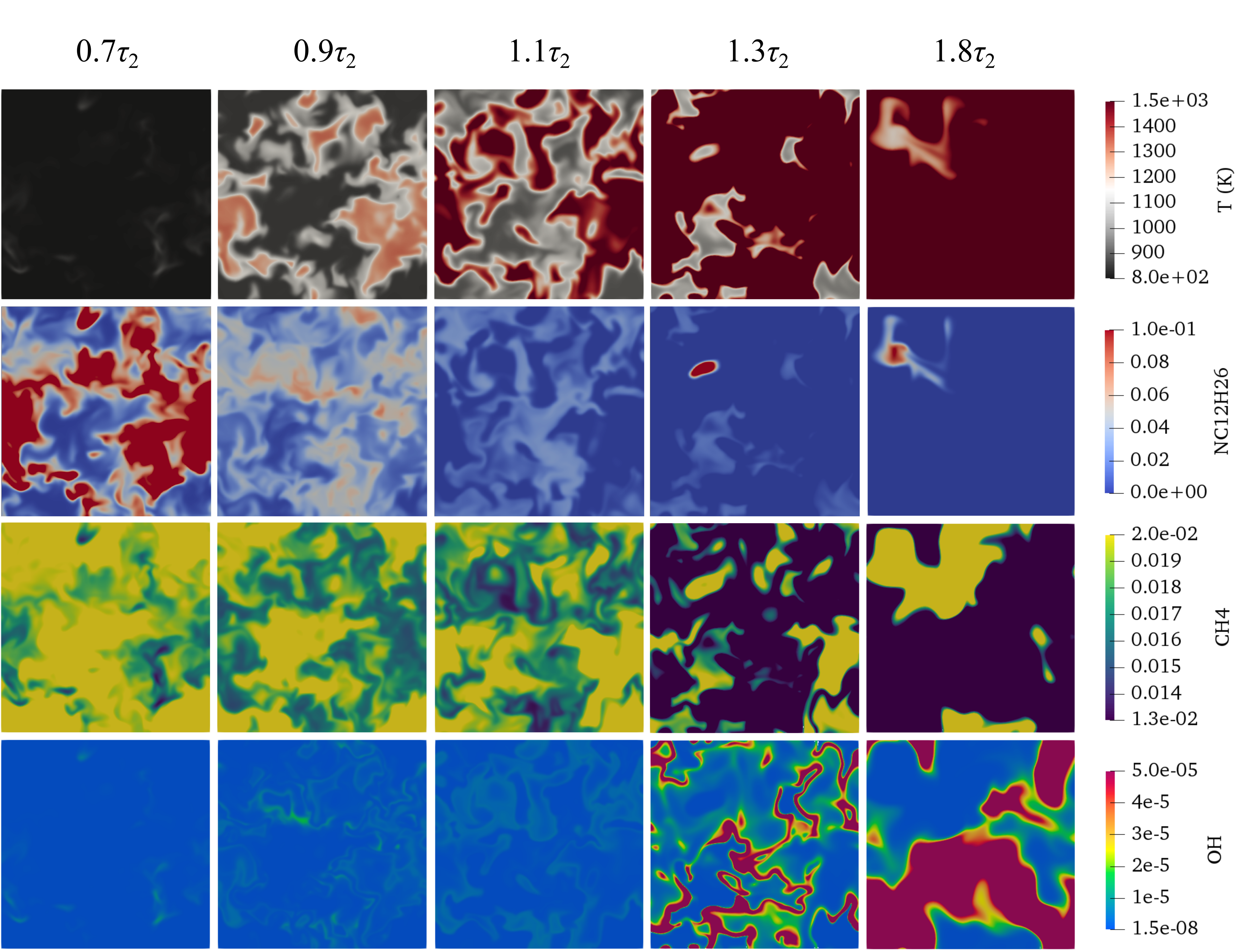}
    \caption{Temporal evolution of temperature (top row), $n$-dodecane (second row), methane (third row), and \ce{OH} (bottom row) are shown in cutplanes taken from the 3D stratified combustion simulation around the second-stage ignition time. With gradual temperature rise, ignition kernels are formed at different locations. Consistently, the fuels ($n$-dodecane and methane) are consumed and deflagrative-like fronts are formed and finally merged.}
    \label{fig:spark}
\end{figure}

After providing speed-up benchmarks along with scaling tests in the previous section, we next present another academic test case in this section, i.e. 3D stratified combustion~(\Cref{fig:spark}). This case can not be even tested for $10$ iterations using \texttt{Standard} within a reasonable computational time.  Hence, the purpose here is mainly to show that it is possible to investigate 3D ignition/combustion problems using \texttt{DLBFoam+pyJac}. This DF test case has close relevance to combustion mode design in modern engines such as RCCI. 

This test case is conceptually similar to the previous 2D setup by Karimkashi et al.~\cite{Karimkashi2020}.  As shown in the schematic \Cref{fig:schematic_3dcomb}, the computational domain is a 3D cube with periodic boundary conditions and a side length $D=\SI{1}{\milli\meter}$. The grid size is \SI{4}{\micro m} with $256$ grid points along each direction, to capture the flame with at least $10$ grid points within the laminar flame thickness ($\delta_f \approx \SI{40}{\micro m}$). The initial pressure is set at \SI{60}{bar} and the initial temperature is homogeneous within the entire computational domain at \SI{800}{K}, which is relevant to the estimated temperature at the most reactive mixture fraction for DF methane/$n$-dodecane based on 0D simulations ~\cite{Karimkashi2020b, Tekgul2020, Kannan2020, Kannan2021}. Here, pure $n$-dodecane is initially constrained in a blob in the middle of the domain with a diameter $d=D/3$. The ambient gas consists of premixed methane, oxidizer and EGR at $\phi_{\ce{CH4}}=0.5$, similar to the previous test case, with the composition reported in \Cref{tab:composition_shear_layer}. As in the previous test case, Yao mechanism is used.

Turbulence is initialized using Taylor-Green-Vortex (TGV) structure, which is generated in a separate non-reactive 3D simulation with an initial velocity $U=\SI{50}{m/s}$ and reference length $L_{\text{ref}}=D/2\pi$. We let the initialized TGV to evolve in time until the gradient of total kinetic energy reaches its peak and then, velocity and pressure fields are mapped to the reactive case as initial fields. The estimated Reynolds number at the start of reacting simulation is $Re \approx 1000$. In this case study, the interactive roles of convection, diffusion and reaction are investigated. First, the initial $n$-dodecane is mixed with the ambient gas at early simulation time instances, long before the second-stage ignition ($\tau_2$). Here, $\tau_2 \approx \SI{0.28}{\milli s}$ is defined as the first time instance when $T > \SI{1500}{K}$. At $\tau_2$, the stratified mixture ignites at the most reactive mixture fractions. The ignition front propagates and finally forms laminar premixed flame fronts, which complete the combustion in the standard deflagration mode. The present observations in 3D are in close agreement with the observations made in our previous study in 2D ~\cite{Karimkashi2020}. 

\Cref{fig:spark} depicts 2D cut-planes for temporal evolution of temperature (top row), $n$-dodecane (second row), methane (third row), and hydroxyl radical \ce{OH} (bottom row) wherein time is normalized by $\tau_2$. It is observed that a stratified mixture of $n$-dodecane and methane/oxidizer is formed before any significant temperature rise in the system ($t=0.7\tau_2$). Closer to the ignition time, hot spots emerge ($t=0.9\tau_2$) and deflagrative-like fronts develop at a moderate rate with various wrinkled surfaces ($t=1.1\tau_2$). Slightly after the ignition time, the fronts evolve ($t=1.3\tau_2$) as clearly observed from the \ce{OH} cutplane. Finally, the fronts merge ($t=1.8\tau_2$) and combustion is almost complete. We note that at $t=1.8\tau_2$, only small fractions of $n$-dodecane and methane are left in the system, which are convected from other iso-surfaces to the displayed cutplane in \Cref{fig:spark}. 

In this case study, a speed-up gain by a factor of \numrange{2}{3} orders of magnitude is observed with \texttt{DLBFoam+pyJac} compared to \texttt{Standard}. Therefore, it would not be possible to benchmark this problem using \texttt{Standard} for $100$ iterations. In particular, we evaluate the speed-up of our model against \texttt{Standard} by restarting the simulation from the first-stage ignition time (\SI{0.2}{\milli s}) and using the same number of processors ($1152$ processors) and identical tolerances. While \texttt{Standard} simulation proceeds only $7$ iterations in $\sim 76000$ seconds of clock time, the respective time is only about $200$ seconds with \texttt{DLBFoam+pyJac}; i.e. speed-up by a factor of $\approx 400$ with \texttt{DLBFoam+pyJac} compared to \texttt{Standard} is achieved. As a final note, the total clock time for simulating this case using \texttt{DLBFoam+pyJac} with $1152$ processors for \SI{0.5}{\milli s} simulation time is $\approx 13$ hours. Hence, \texttt{DLBFoam+pyJac} is a key enabler to study such 3D reactive flow configurations with OpenFOAM. 

\subsection{ECN Spray A}

At this point, it is clear that \texttt{DLBFoam+pyJac} offers great advantages over \texttt{Standard}. In particular, \texttt{DLBFoam+pyJac} was shown to scale almost linearly with the number of processors in an academic case and enable 3D reactive simulations with a relatively large chemical kinetic mechanism and tight solver tolerances. Next, we move on to a more realistic 3D configuration i.e. the Engine Combustion Network Spray A test case with experimental validation data. Spray A represents an igniting non-premixed diffusion flame under engine-relevant conditions. Hence, it is an optimal test case for reactive CFD large-eddy simulation (LES) code validation.

\begin{figure}[H]
    \centering
    \includegraphics[width=0.98\textwidth]{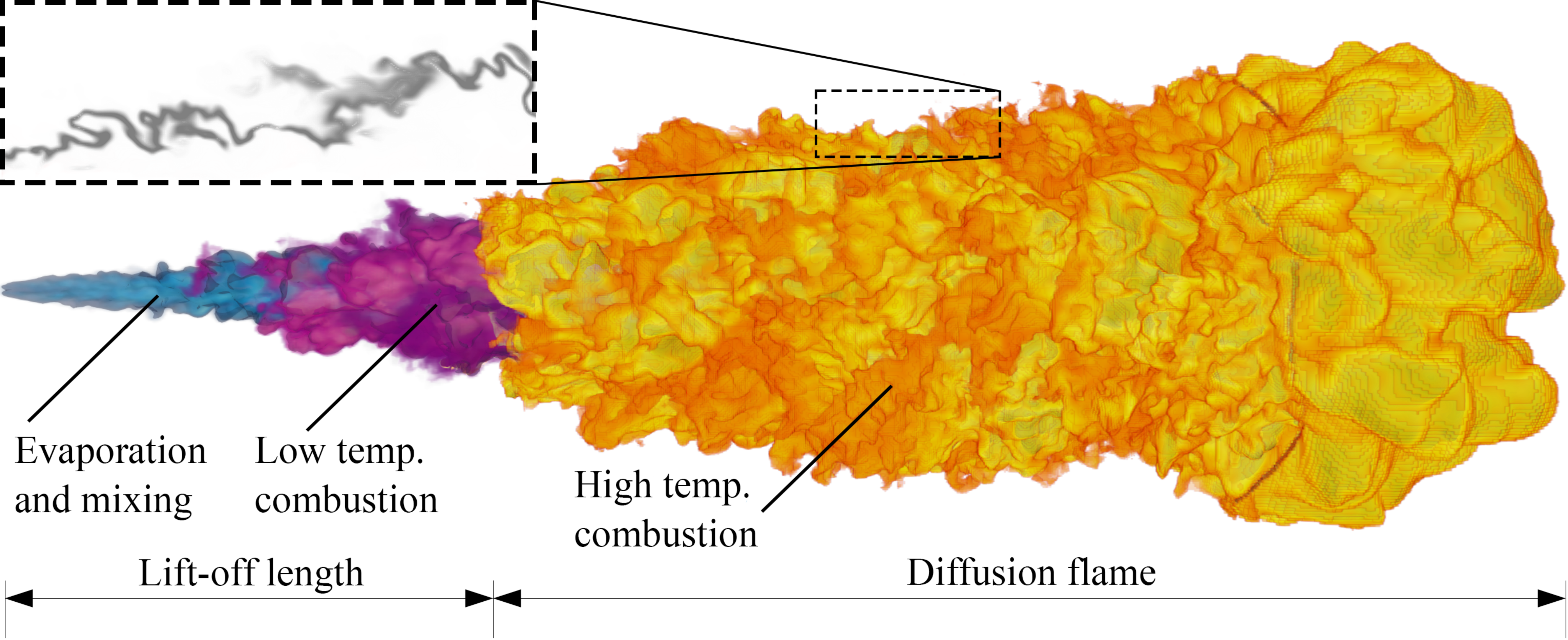}
    \caption{Volume rendering of the ECN Spray A diffusion flame of ambient $T=900$~\si{K} after \SI{1.5}{ms}. The inset in the top left corner presents a 2D projected cut-plane of the \ce{OH} radical mass fraction.}
    \label{fig:volRend}
\end{figure}

A schematic of the computational setup is demonstrated in \Cref{fig:schematic_sprayA}, where the domain volume corresponds to that of ECN combustion vessel by Sandia. Liquid $n$-dodecane at temperature of \SI{363}{K} and pressure of \SI{150}{MPa} is injected from a \SI{90}{\micro m} nozzle (i.e. Spray A) into a hot quiescent ambient gas of density \SI{22.8}{kg/m^3} and mixture composition (\SI{15}{\percent}~\ce{O2}, \SI{75.15}{\percent}~\ce{N2}, \SI{6.23}{\percent}~\ce{CO2} and \SI{3.62}{\percent}~\ce{H2O}) based on the molar fractions. The simulations are performed for $4$ reacting cases with varied ambient temperature  ($900$, $1000$, $1100$, and $1200$~\si{K}), and a single non-reacting case at \SI{900}{K}. The domain is discretized with static mesh of \SI{62}{\micro m} cell size in the inner most refinement layer to resolve the turbulent mixing and the quasi-steady lifted flame, as shown in \Cref{fig:volRend}, resulting in total amount of $25$--$39$ million cells depending on the particular case. Moreover, the implicit LES (ILES) approach is employed for turbulent subgrid-scale (SGS) modeling, consistent with our previous spray combustion works~\cite{Tekgul2021a, Gadalla2021b, Kannan2021, Gadalla2020, Kannan2020, Tekgul2020, Antonelli2011}, while the no-breakup model approach is used for droplet atomization which is further discussed by Gadalla et al.~\cite{Gadalla2020}. The volume-rendered spray flame shown in \Cref{fig:volRend} for ambient $T=900$~\si{K} indicates the mixing and evaporation zone, the low-temperature combustion zone, and the ignition and high temperature combustion region wherein fully developed diffusion flame is observed.

\begin{figure}[H]
    \centering
    \subfloat[][]{\includegraphics[width=0.450\textwidth]{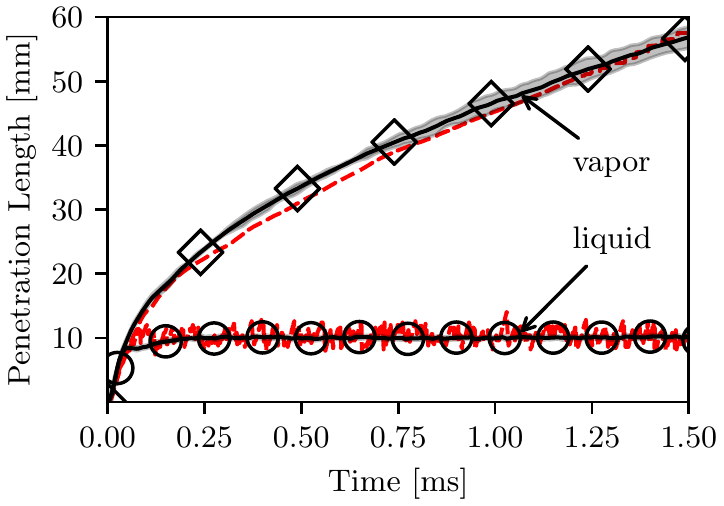}\label{fig:penet}}%
    \hspace{1.5mm}
    \subfloat[][]{\includegraphics[width=0.450\textwidth]{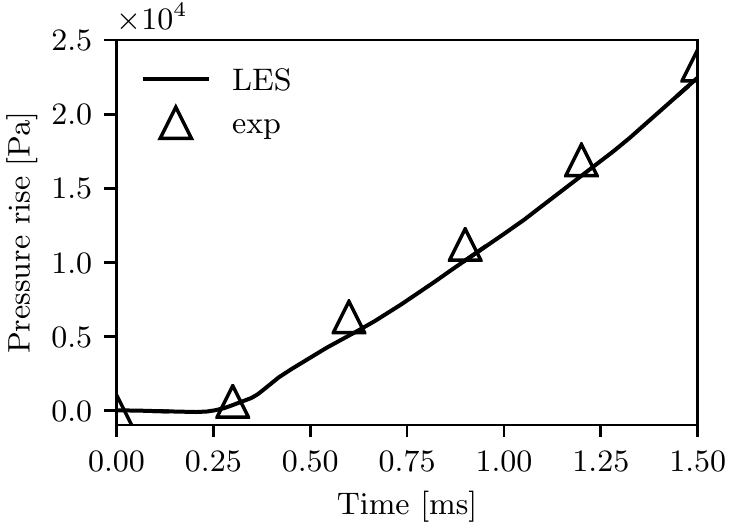}\label{fig:prise}}%
    
    \vspace*{-4mm}
    
    \subfloat[][]{
    \begin{minipage}{0.450\textwidth}
	  \includegraphics[width=\textwidth]{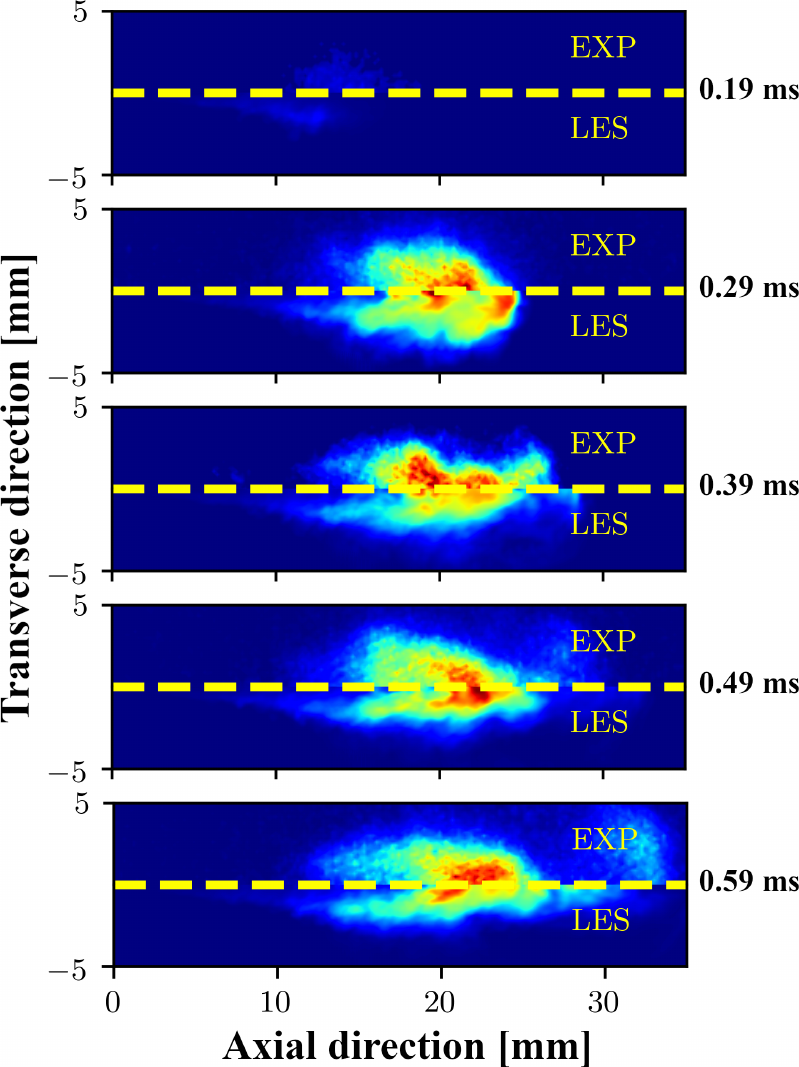}
    \end{minipage}
    \label{fig:ch2o}
    }
    \subfloat[][]{
    \begin{minipage}{0.450\textwidth}
	    \includegraphics[width=\textwidth]{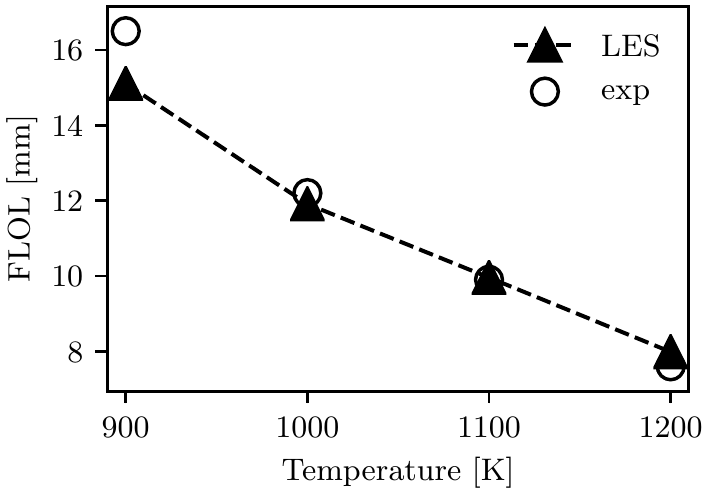}
	    \includegraphics[width=\textwidth]{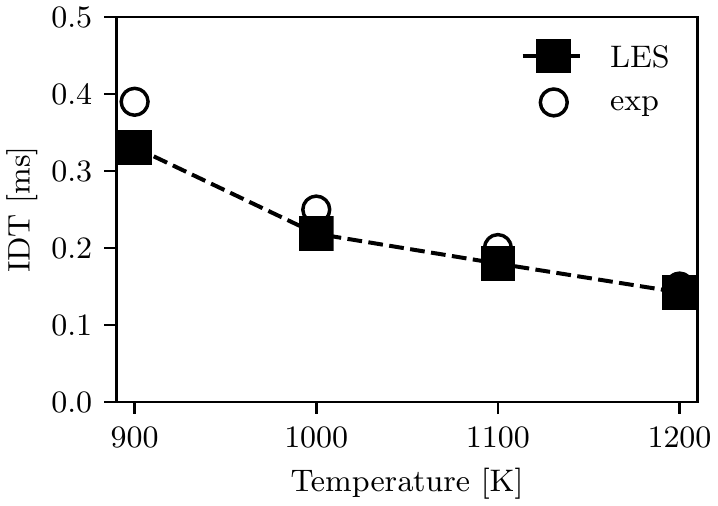}
      \end{minipage}
    \label{fig:Tsweep}
    }
    \caption{Validations of the ECN Spray A at ambient $T=900$~\si{K} for (a) liquid and vapor penetrations under non-reactive conditions, (b) pressure rise in the combustion vessel after ignition, (c) formaldehyde fields at various snapshots, and (d) ignition delay time and flame lift-off length (FLOL) at varied ambient temperature levels of the mixture.}
    \label{fig:spray_a}
\end{figure}

The present show case demonstrates the capability of our new model to match the well-known ECN Spray A benchmark. First, in \Cref{fig:penet} liquid and vapor penetrations are validated against experimental data up until \SI{1.5}{ms} under non-reacting conditions at \SI{900}{K}. Then, the simulation is repeated in reacting mode and the pressure rise is monitored in \Cref{fig:prise} against experiments for the same temporal interval. In \Cref{fig:ch2o} the formaldehyde (\ce{CH2O}) planar laser-induced fluorescence (PLIF) false color images obtained by Skeen et al.~\cite{Skeen2015} at various snapshots are compared against the corresponding LES data using the new solver. Here, the LES data are circumstantially averaged for each sampling point in the axial-radial plane, which are then compared with the experimental data set. This comparison verifies the global trends for Spray A baseline including ignition onset, cool flame, and quasi-steady flame lift-off. Finally, validations of ignition delay times (IDTs) and flame lift-off lengths against experimental data at various temperature levels are presented in \Cref{fig:Tsweep}.

\begin{figure}[h]
    \centering
    \includegraphics[width=0.47\textwidth]{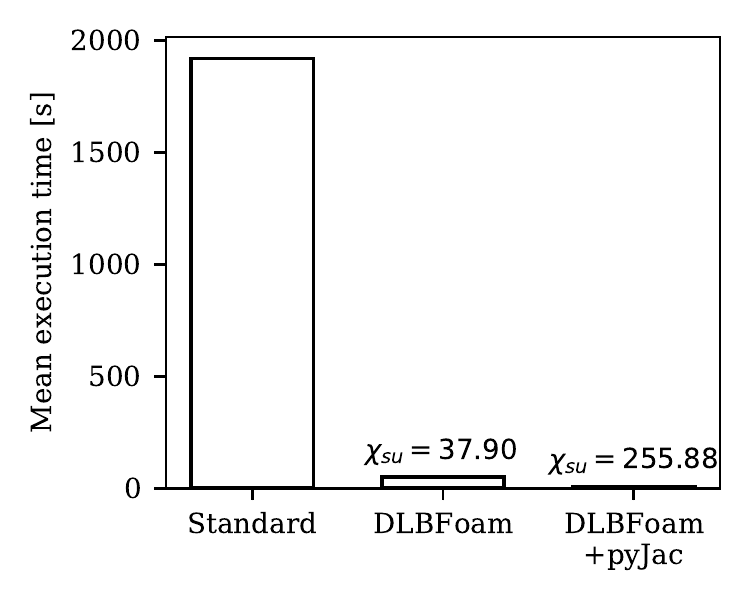}
    \caption{Mean execution times over $65$ iterations for the ECN Spray A after \SI{0.3}{ms} of the simulation. $\chi_{su}$ is the speed-up factor of \texttt{DLBFoam} or \texttt{DLBFoam+pyJac} compared to \texttt{Standard}.}
    \label{fig:speedup_constDt}
\end{figure}

After that, we demonstrate in \Cref{fig:speedup_constDt} the gained speed-up using \texttt{DLBFoam+pyJac} within $100$ time steps after \SI{0.3}{ms} of the simulation and using a constant step size of \SI{2e-7}{s}. The load imbalance in this test case is rather high. It originates from the large number of ambient cells which are computationally less stiff than those in the ignition region. This imbalance is further increased with reference mapping so that ambient cells become mostly idle, hence improving the performance gain of dynamic load balancing, which results in a factor of $38$ speed-up, compared with \texttt{Standard}. After that, a further speed-up by a factor of $6.7$ is attributed to the analytical Jacobian retrieval and the robust linear algebra, i.e. \texttt{DLBFoam+pyJac}, hence a total speed-up by a factor of $256$ as compared to \texttt{Standard}.

\subsection{Sandia flame D}

The ECN Spray A case demonstrated excellent performance for LES of  non-premixed spray flame using \texttt{DLBFoam+pyJac}. The last test case is the Sandia flame D~\cite{Barlow1998} representing a piloted partially premixed methane-air flame. Sandia flame D is a well-known configuration from the TNF workshop flame series, which is often studied in the literature, e.g.~\cite{Pitsch2000, Vreman2008, Bertels2019, Garmory2013}. Therefore, it is considered as a useful benchmark for reactive CFD LES code validation for gas burner flame investigations.

\begin{figure}[H]
    \centering
        \begin{minipage}{0.42\textwidth}
    	    \includegraphics[width=\textwidth]{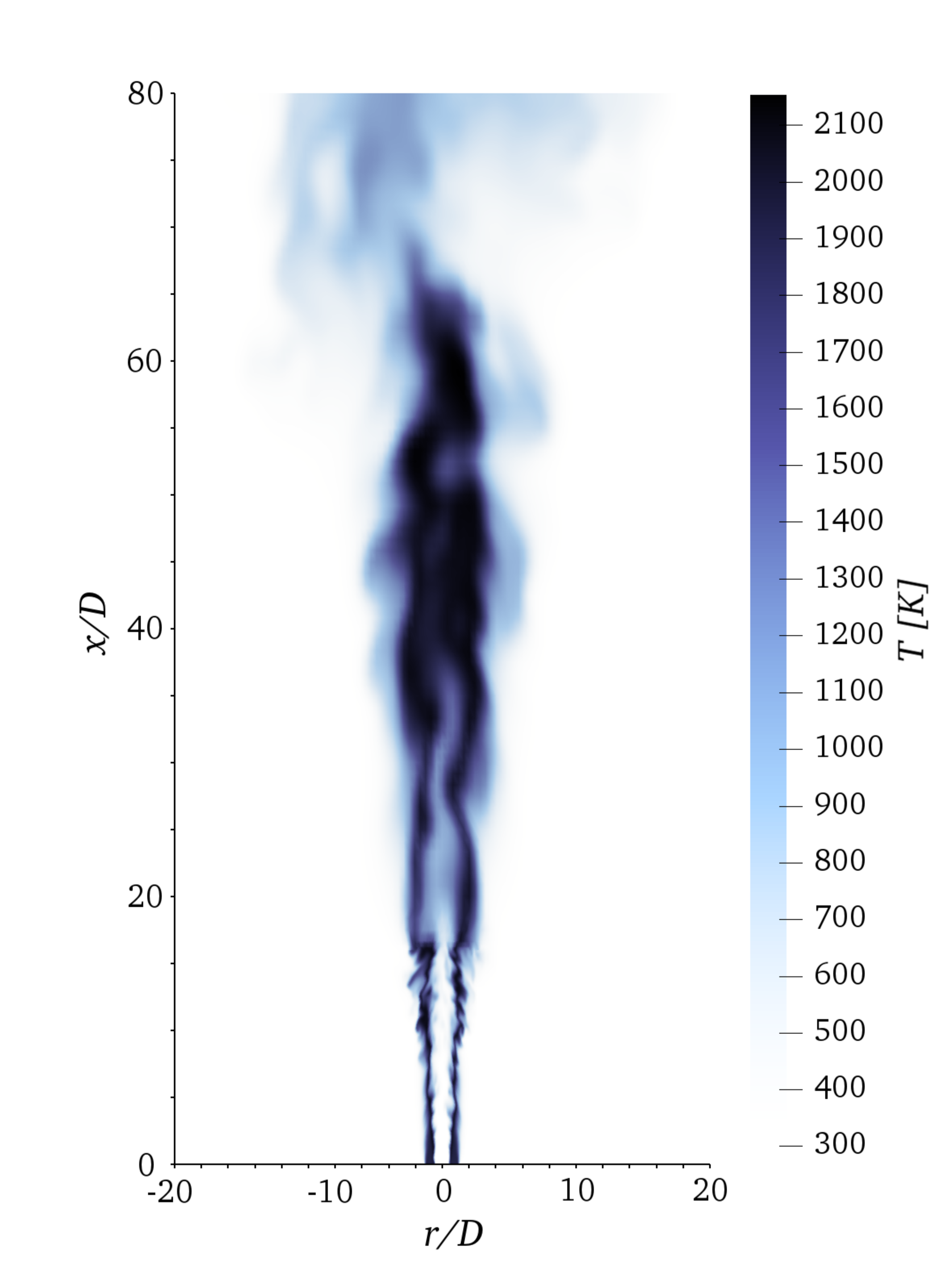}
          \end{minipage}
    \label{fig:sandiad_Tfield}
    \
        \begin{minipage}{0.42\textwidth}
    	  \includegraphics[width=\textwidth]{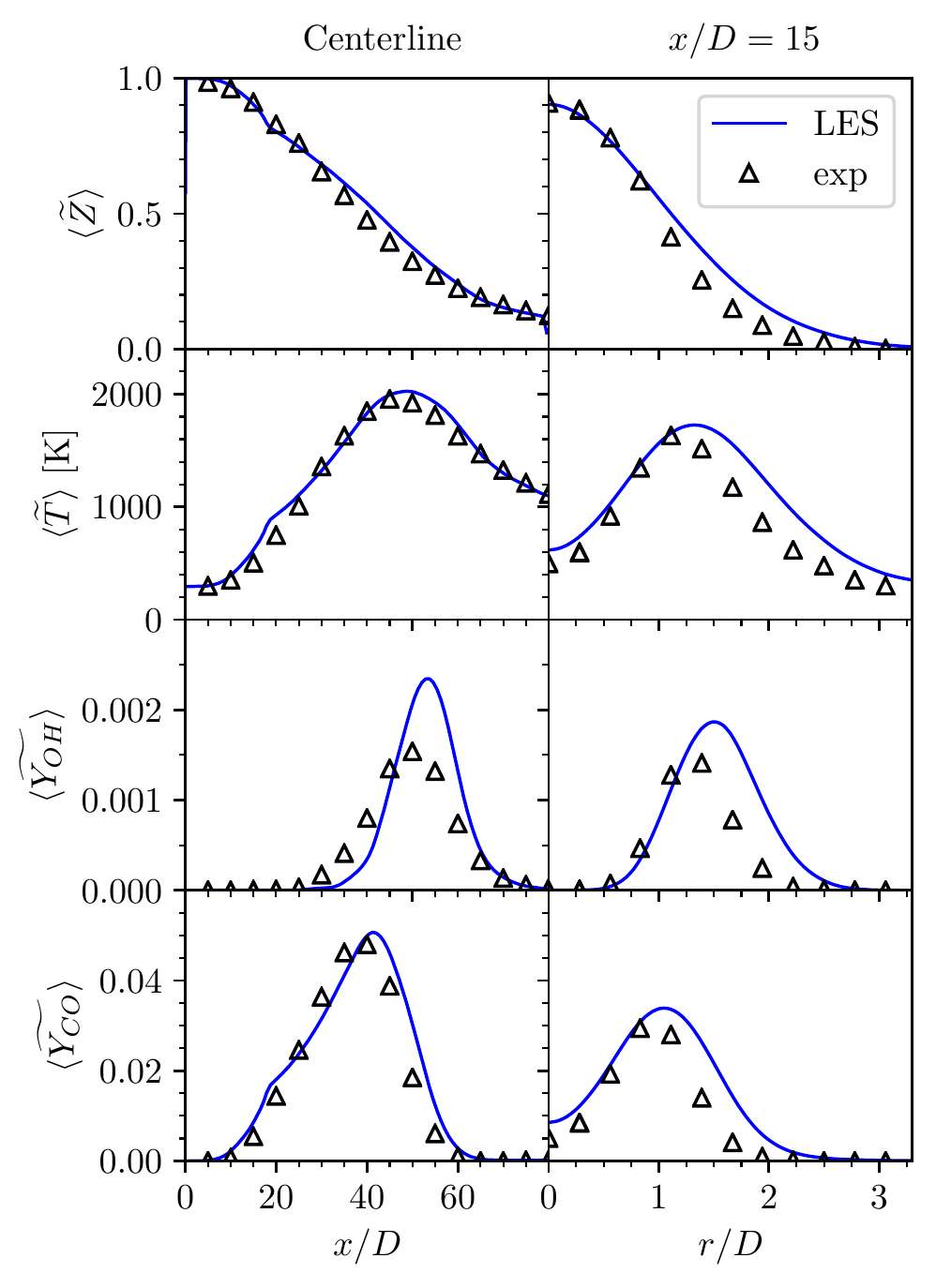}
        \end{minipage}
    \label{fig:sandiad_validation}
    
    \caption{Top: Instantaneous temperature field of Sandia flame D at $t=\SI{0.1}{s}$. Note that the data attributed to the refined grid is only available to $x/D = 15$ in order to limit the computational cost. Bottom: Validations of the Sandia flame D for mean mixture fraction, temperature and species mass fraction at centerline and $x/D=15$.}
    \label{fig:sandiad_results}
\end{figure}

A schematic of the Sandia flame D setup is provided in \Cref{fig:schematic_sandiaD}. The main jet is a mixture of \ce{CH4} and air at equivalence ratio $\phi_{\ce{CH4}} = 3.17$. It is injected with a bulk velocity of \SI{49.6}{m/s} and temperature of \SI{294}{K} to the domain from a nozzle of diameter $D=\SI{7.2}{mm}$. The flame is stabilized with a pilot jet which is a hot (\SI{1880}{K}) lean mixture of \ce{C2H2}, \ce{H2}, air, \ce{CO2} and \ce{N2}. The inner and outer diameters of the pilot nozzle are \SI{7.7}{mm} and \SI{18.2}{mm}, respectively, and the flow velocity is \SI{11.4}{m/s}. Jet inlets are surrounded by air co-flow with an inner diameter of \SI{18.9}{mm}. A~Cartesian mesh is used, and a refinement zone with cell size of \SI{0.75 x 0.368 x 0.368}{mm} defined from the nozzle exit until $15D$ is realized, resulting in a total of $2M$ cell count. Velocity fluctuations in the form proposed by Pitsch and Steiner~\cite{Pitsch2000} are superimposed on the main jet inlet velocity profile to induce turbulent jet behaviour. Here, the purpose is to demonstrate that the code produces first order statistics for this flame configuration. Hence, we do not focus on the turbulence-chemistry interaction modeling but use the ILES approach as a SGS scale model, consistent to the previous test cases. A reduced chemical kinetic mechanism DRM19~\cite{DRM19} with $21$ species and $83$ reactions is used. The simulation is performed for \SI{100}{ms} after which the fields are averaged for another \SI{150}{ms}, where the flame becomes statistically stable.

\Cref{fig:sandiad_results} depicts time-averaged mixture fraction, temperature and  specie mass fractions profiles compared to the experimental data~\cite{Barlow1998} and an instantaneous snapshot of the temperature field. The close agreement between experiments and predicted mixture fraction and temperature profiles along the center axis indicates that the coupled influence of turbulent mixing and flame heat release is resolved adequately in the core of the jet flame. Minor species profiles, \ce{CO} and \ce{OH}, also correspond well with the experiments. Predicted \ce{OH} mass fractions show correct trends, however the maximum value is overpredicted compared with the measured data. Overall, mass fraction prediction of short lived species like \ce{OH} is a challenging task due to the nonlinear evolution of the species~\cite{Vreman2008,Bertels2019,Garmory2013, Lysenko2014} and it is out of the scope of the present study. In the radial direction, the reactions are mostly occurring in the shear layer between the main jet and co-flow ($r/D = 1$), which is evident in the peaks of the mass fraction profiles. The radial profiles agree well with the experimental data on the rich side. However, mixing is overpredicted in the lean region ($r/D > 1$). Close to the nozzle and away from the central axis, the reaction zone may still be considered relatively laminar~\cite{Pitsch2000}. Therefore, the overpredicted profiles in the lean region could be explained by the utilized unity Lewis number assumption. Moreover, the temperature and species profiles also follow the same trend as the mixture fraction in the flame on the rich and lean sides.

\begin{figure}[H]
    \centering
    \includegraphics[width=0.47\textwidth]{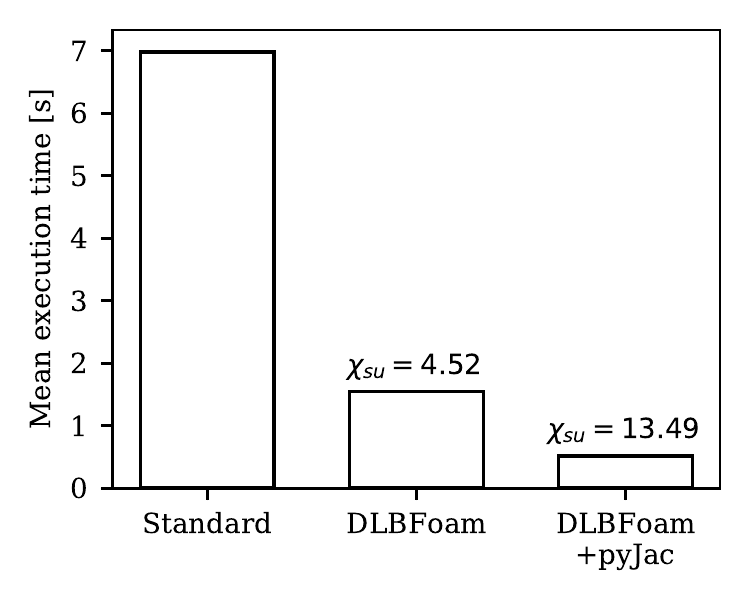} \\
    \caption{
    Mean execution times over $100$ iterations for the Sandia flame D after \SI{0.1}{s} of the simulation. $\chi_{su}$ is the speed-up factor of \texttt{DLBFoam} or \texttt{DLBFoam+pyJac} compared to \texttt{Standard}.}
    \label{fig:sandiad_benchmark}
\end{figure}

Next, a speed-up test is carried out for a stabilized flame for $100$ iterations with a constant time step of \SI{1e-6}{s}. The results are presented in \Cref{fig:sandiad_benchmark}. The \texttt{DLBFoam} provides speed-up by a factor of $4.52$, which is significantly lower than with ECN Spray A. This can be explained by a relatively low number of mapped cells due to the location of the refinement zone in the core of the jet. Optimized ODE solver routines \texttt{pyJac} and \texttt{LAPACK} bring additional speed-up by a factor of $2.99$. Here, the lower speed-up is a consequence of the use of the reduced mechanism, which as discussed earlier significantly reduces computational cost of the chemistry problem. However, the combined effect of the \texttt{DLBFoam+pyJac} is significant and brings about $1$ order of magnitude speed-up to this case.

\section{Conclusions}
\label{sec:conc}
In this work, an open-source dynamic load balancing library for OpenFOAM, namely \texttt{DLBFoam}, with reference mapping feature was re-introduced and improved. In particular, the feature was extended with optimized chemistry ODE solution routines. First, the Jacobian evaluation procedures in OpenFOAM were replaced with an analytical implementation provided by the open-source package \texttt{pyJac}. This step resulted in one order of magnitude speed-up in a 0D homogeneous reactor test case. Then in order to fully utilize the benefits of the new fully dense Jacobian, LU decomposition and back substitution routines were replaced with more robust ones from the open-source library \texttt{LAPACK}, resulting in even higher speed-up especially for tighter ODE tolerances.

The improvements were tested in two academic cases and then applied to two experimental flame cases. First, the model was tested in a 2D reacting shear layer problem, demonstrating almost perfectly linear scaling. Speed-up by a factor of $\approx 30$ was demonstrated in the 2D case for \texttt{DLBFoam} with \texttt{pyJac} and \texttt{LAPACK}. Second, 3D stratified combustion was studied, which was not even computationally feasible to benchmark with the \texttt{Standard} model due to $2$--$3$ orders of magnitude performance difference. Last, two 3D flames were investigated: ECN Spray A and Sandia Flame D showing speedup factors of $\approx 256$ and  $\approx 13.5$ respectively for \texttt{DLBFoam} with \texttt{pyJac} and \texttt{LAPACK}. We also note that the reported speed-up numbers in comparison to \texttt{Standard} were achieved with fixed ODE tolerances, specific time step sizes, and chemical kinetic mechanisms. It is worth acknowledging that changing those parameters may affect the speed-up estimates. However, the promising results herein indicate that the improvements to \texttt{DLBFoam} offer an avenue to model complex combustion phenomena in OpenFOAM with increased accuracy and computational efficiency. Further investigations in a broader user community could result in more in-depth analysis and development of the present test cases for different combustion models and chemical mechanisms. As future work, more test cases could be established around the proposed reactive CFD approach to account for a broader variety of turbulent combustion conditions. Finally, the source code and the case setup files used in this work are openly available~\cite{DLBFoam_repo, DLBFoam_cases_repo}

\section*{Acknowledgements}

This study is financially supported by the Academy of Finland (grant numbers 318024, 332784, and 332835). 
Computational resources have been provided by CSC -- Finnish IT Center for science and Aalto Science-IT project. 
We gratefully thank Dr. Heikki Kahila (W\"{a}rtsil\"{a} Finland Oy) and Dr. Petteri Peltonen (VTT Oy) for their precious comments and technical advice throughout the implementation process.

\appendix

\section{Supplementary material}
\label{sec:append_A}
Library source code is openly available in a public GitHub repository~\cite{DLBFoam_repo}. Test case setups are openly available in a separate public GitHub repository~\cite{DLBFoam_cases_repo}.

\bibliography{biblio.bib}
\bibliographystyle{elsarticle-num.bst}

\end{document}